\documentclass[pre,superscriptaddress,twocolumn]{revtex4}
\pdfoutput=1
\usepackage[colorlinks=true,urlcolor=blue]{hyperref}
\usepackage{amsfonts}
\usepackage{graphicx}
\usepackage{amsmath}
\usepackage{xcolor}
\usepackage{bm}

\usepackage{color}
\definecolor{red}{rgb}{0.75,0,0}
\definecolor{blue}{rgb}{0,0,0.75}
\definecolor{green}{rgb}{0,0.5,0}

\newcommand{\arxivhref}[2]{{\hypersetup{urlcolor=green}\href{#1}{#2}}}  
\newcommand{\revision}[1]{#1}

\DeclareMathOperator{\tr}{tr}
\DeclareMathOperator{\erf}{erf}
\DeclareMathOperator{\erfc}{erfc}

\begin{document}

\title{Geometry and topology of turbulence in active nematics}

\author{Luca Giomi}
\affiliation{Instituut-Lorentz, Universiteit Leiden, P.O. Box 9506, 2300 RA Leiden, The Netherlands}

\begin{abstract}
The problem of low Reynolds number turbulence in active nematic fluids is theoretically addressed. Using numerical simulations I demonstrate that an incompressible turbulent flow, in two-dimensional active nematics, consists of an ensemble of vortices whose areas are exponentially distributed within a range of scales. Building on this evidence, I construct a mean-field theory of active turbulence by which several measurable quantities, including the spectral densities and the correlation functions, can be analytically calculated. Because of the profound connection between the flow geometry and the topological properties of the nematic director, the theory sheds light on the mechanisms leading to the proliferation of topological defects in active nematics and provides a number of testable predictions. A hypothesis, inspired by Onsager's statistical hydrodynamics, is finally introduced to account for the equilibrium probability distribution of the vortex sizes. 
\end{abstract}

\maketitle

The paradigm of ``active matter'' \cite{Ramaswamy:2010,Vicsek:2012,Marchetti:2013} has had notable successes over the past decade in describing self-organization in a surprisingly broad class of biological and bio-inspired systems: from flocks of starlings \cite{Vicsek:1995,Ballerini:2008} to robots \cite{Giomi:2013a}, down to bacterial colonies \cite{Dombrowski:2004,Wogelmuth:2008,Zhang:2010,Wensink:2012,Dunkel:2013}, motile colloids \cite{Bricard:2013,Palacci:2013} and the cell cytoskeleton \cite{Kruse:2004,Voituriez:2005,Kruse:2005}. Active systems are generic non-equilibrium assemblies of anisotropic components that are able to convert stored or ambient energy into motion. Because of the interplay between internal activity and the interactions between the constituents, these systems exhibit a spectacular variety of collective behaviors that are entirely self-driven and do not require a central control mechanism. 

A particularly interesting manifestation of collective behavior in active systems is the emergence of spatio-temporal chaos. In active bio-fluids, such as bacterial suspensions or cytoskeletal mixtures, the chaotic dynamics takes place through the formation of structures, such as jets or swirls, reminiscent of turbulence in Newtonian fluids, in spite of the undisputed predominance of dissipation over inertia at the microscopic scale. Examples of low Reynolds number turbulence in active fluids were first reported for the case of bacterial suspensions \cite{Dombrowski:2004,Wogelmuth:2008,Zhang:2010,Wensink:2012,Dunkel:2013}, where this is believed to have an important impact on nutrient mixing and molecular transport at the microbiological scale. 

\begin{figure}[t]
\centering
\includegraphics[width=1\columnwidth]{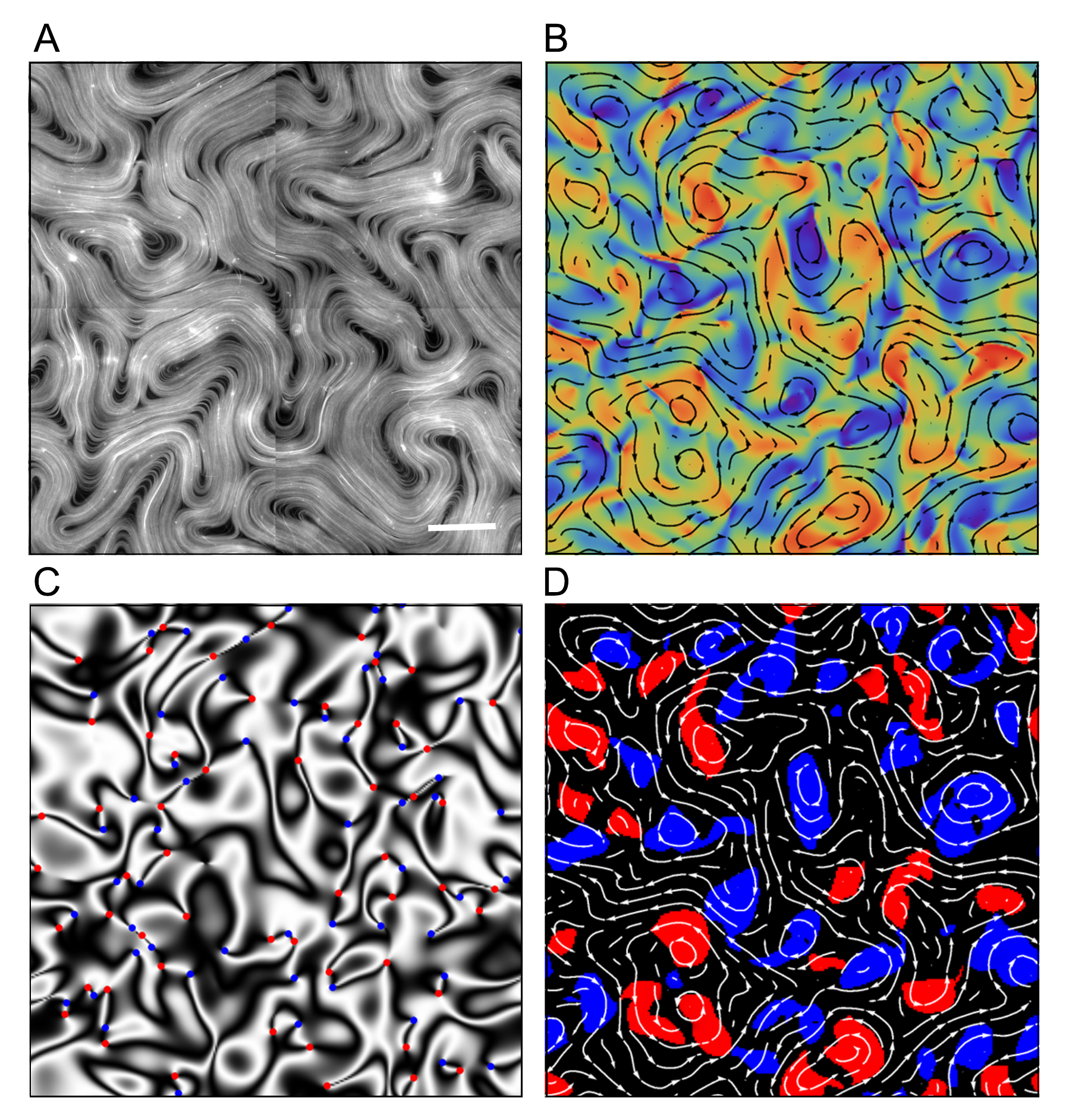}	
\caption{\label{fig:snapshots}({A}) A two-dimensional active nematic suspensions of microtubules bundles and kinesin at the water-oil interface. The white scale bar corresponds to 100 $\mu$m. [Courtesy of the Dogic Lab]. ({B}-{D}) Numerical simulations of an extensile active nematic obtained from and integration of Eqs. \eqref{eq:hydrodynamics}. ({B}) Flow velocity (black streamlines) and vorticity (background color). ({C}) {\em Schlieren} texture constructed from the director field $\bm{n}$. The red and blue dots mark respectively the $+1/2$ and $-1/2$ disclinations. ({D}) Clockwise rotating (blue) and counterclockwise rotating (red) vortices, detected by measuring the Okubo-Weiss field as described in the text.}
\end{figure}

Recently, a series of remarkable experiments on actomyosin motility assays \cite{Schaller:2013} and suspensions of microtubles bundles and kinesin \cite{Sanchez:2012,Keber:2014} (see Fig. \ref{fig:snapshots}A), have unveiled a profound link between the topological structure of the orientationally ordered constituents and the flow dynamics, suggesting that active turbulence could be mediated by unbound pairs of topological defects. According to this picture, the strong distortion associated with a defect, fueled by the active stresses, determines a local shear flow which in turn drives the unbinding of more defect pairs. Similar patterns have been observed in ``living liquid crystals'' obtained from the combination of swimming bacteria and lyotropic liquid crystals \cite{Zhou:2014}. Whether these examples of {\em active turbulence} are different realizations of the same universal mechanism or substantially different forms of spatio-temporal chaos, represents a profound and yet unsolved problem.

The current efforts toward understanding active turbulence rely on the use of continuum models, such as the Toner-Tu or Swift-Hohenberg model \cite{Wensink:2012,Dunkel:2013} or the equations of active nematodynamics \cite{Thampi:2013,Thampi:2014a,Thampi:2014b,Giomi:2013b,Giomi:2014}. Both of these approaches have been shown to be able to account for the occurrence of self-sustained low Reynolds number turbulence such as that observed in the experiments on bacteria and cytoskeletal fluids, although a systematic comparison between theory and experiments is still in its infancy. The recent numerical work by Thampi \emph{et al}. \cite{Thampi:2013,Thampi:2014a,Thampi:2014b}, in particular, has provided a convincing demonstration of the correlation between defects dynamics and turbulence in active nematics. The interplay between defects and turbulence has been further investigated in Ref. \cite{Giomi:2013b,Giomi:2014} and, following a different approach, in Ref. \cite{Gao:2014}. Agent-based simulations have also been recently employed to highlight the interplay between defects and dynamics in granular active nematics \cite{Shi:2013}. The overlap between these ``dry'' systems and active liquid crystals remains, however, unclear.

In this article I report an exhaustive numerical study of turbulence in active nematics. As a starting point, I demonstrate that, as for inertial turbulence, low Reynolds number turbulence in active nematics is, in fact, a multiscale phenomenon characterized by the formation of vortices spanning a range of length scales. Within this \emph{active range}, the areas of the vortices are exponentially distributed, while their vorticity is approximately constant. Building on these observations, I then formulate a mean-field theory of turbulence in active nematics that allows the analytical calculation of several measurable quantities, including the mean kinetic energy and enstrophy, their corresponding spectral densities and the velocity and vorticity correlation functions. The connection between the topological structure of the nematic phase and the geometry of the flow is then elucidated through a quantitative description of the defect statistics. 

\section{Results}

\subsection{Active nematdynamics}

Let us consider an incompressible uniaxial active nematic liquid crystal in two spatial dimensions. The two-dimensional setting is appropriate to describe experiments such as that by Sanchez {\em et al}. \cite{Sanchez:2012}, where the microtubule bundles are confined to a water-oil interface forming a dense active nematic monolayer, but also of considerable interest in its own right. Let then $\rho$ and $\bm{v}$ be the density and velocity of an incompressible nematic fluid. Incompressibility requires $\nabla\cdot\bm{v}=0$. Nematic order is described by the alignment tensor $Q_{ij}=S(n_{i}n_{j}-\delta_{ij}/2)$, with $\bm{n}$ the director and $0\le S \le 1$ the nematic order parameter. The tensor $Q_{ij}$ is by construction traceless and symmetric and has only two independent components in two dimensions. The hydrodynamic equations of an active nematic can be constructed from phenomenological arguments \cite{Kruse:2004,Marenduzzo:2007,Giomi:2012}, or derived from microscopic models \cite{Liverpool:2008,Lau:2009} in the form:
\begin{subequations}\label{eq:hydrodynamics}
\begin{gather}
\rho\frac{D v_{i}}{Dt} = \eta\nabla^{2}v_{i}-\partial_{i}p+\partial_{j}\sigma_{ij}\;,\\
\frac{D Q_{ij}}{Dt} = \lambda S u_{ij}+Q_{ik}\omega_{kj}-\omega_{ik}Q_{kj}+\gamma^{-1}H_{ij}\;.	
\end{gather}
\end{subequations}
Here $D/Dt=\partial_{t}+\bm{v}\cdot\nabla$ indicates the material derivative, $p$ is the pressure, $\eta$ the shear viscosity, $\lambda$ the flow alignment parameter and $\gamma$ the rotational viscosity \cite{DeGennes:1993}. In Eq. (\ref{eq:hydrodynamics}b) $u_{ij}=(\partial_{i}v_{j}+\partial_{j}v_{i})/2$ and $\omega_{ij}=(\partial_{i}v_{j}-\partial_{j}v_{i})/2$, are the strain rate and vorticity tensors corresponding to the symmetric and antisymmetric parts of the velocity gradient, while $H_{ij}=-\delta F_{\rm LdG}/\delta Q_{ij}$ is the so-called molecular tensor, governing the relaxational dynamics of the nematic phase and obtained from the two-dimensional Landau-de Gennes free energy \cite{DeGennes:1993}:
\begin{equation}\label{eq:landau_degennes}
F_{\rm LdG} = \frac{1}{2}\int d^{2}r\,\left[K|\nabla\bm{Q}|^{2}+C\tr\bm{Q}^{2}(\tr\bm{Q}^{2}-1)\right]\;,	
\end{equation}
with $K$ and $C$ material constants. Finally, the stress tensor $\sigma_{ij}=\sigma_{ij}^{\rm e}+\sigma_{ij}^{\rm a}$ is the sum of the elastic stress $\sigma_{ij}^{\rm e}=-\lambda H_{ij}+Q_{ik}H_{kj}-H_{ik}Q_{kj}$, due to the entropic elasticity of the nematic phase, and an active contribution $\sigma_{ij}^{\rm a}=\alpha Q_{ij}$ describing the contractile $(\alpha>0)$ and extensile $(\alpha<0)$ stresses exerted by the active particles in the direction of the director field. \revision{The Ericksen stress $\sigma_{ij}^{\rm E}=-\partial_{i}Q_{kl}\,\delta F_{\rm LdG}/\delta(\partial_{j}Q_{kl})$ has been neglected because of higher order in the derivatives of $Q_{ij}$ compared to $\sigma_{ij}^{\rm e}$. This simplification is known not to have appreciable consequences in the fluid mechanics of two-dimensional active nematics \cite{Marenduzzo:2007,Giomi:2012}}.

Eqs.~\eqref{eq:hydrodynamics} have been numerically integrated in a square domain of size $L$ with periodic boundary conditions.  To render the equations dimensionless, all the variables have been normalized by the typical scales associated with the viscous flow. Distances are then scaled by the system size $L$, time by the time scale of viscous dissipation $\tau=\rho L^{2}/\eta$ and stress by the viscous stress scale $\Sigma=\eta/\tau$. Finally, low Reynolds number is imposed by setting $Dv_{i}/Dt=\partial_{t}v_{i}$ in Eq. (\ref{eq:hydrodynamics}a). The integration is performed by finite differences on a square grid of $256\times 256$ points via a fourth-order Runge-Kutta method. To make contact with the recent and ongoing experiments on microtubule suspensions \cite{Sanchez:2012,Keber:2014}, I restrict the discussion to the case of extensile systems ($\alpha<0$). The contractile case was found to be nearly identical and is briefly described in Appendix \ref{sec:extensile_vs_contractile}. Unless stated otherwise the parameter values used in the numerical simulations are $\lambda=0.1$, $K=1$, $\gamma=10$ and $C=4\times 10^{4}$, in the previously described units.    

\subsection{\label{sec:active_range}Active range}

Eqs. \eqref{eq:hydrodynamics} contain two important length scales, in addition to the system size $L$. These are the coherence length of the nematic phase $\ell_{\rm n}=\sqrt{K/C}$ and the active length scale $\ell_{\rm a}=\sqrt{K/|\alpha|}$. The former determines how quickly the nematic order parameter drops in the neighborhood of a topological defect and can be taken as a measure of the defect core radius. The quantity $\ell_{\rm a}$, on the other hand, is the length scale over which active and passive stresses balance, leading to spontaneous elastic distortion and hydrodynamic flow \cite{Thampi:2014a,Thampi:2014b,Giomi:2014}. As a consequence, a quiescent uniformly oriented configuration becomes unstable to a laminar flowing state once $\ell_{\rm a}\sim L$ \cite{Voituriez:2005,Marenduzzo:2007,Giomi:2012,Edwards:2009}. As $\ell_{\rm a}$ becomes lower than the system size $L$, the laminar flow is eventually replaced by a turbulent flow. Depending on the values of the various parameters in Eqs. \eqref{eq:hydrodynamics}, the onset of turbulence can be characterized by the formation of ``walls'', narrow regions where the director is highly distorted, and their breakup into pairs of $\pm 1/2$ disclinations \cite{Thampi:2013,Giomi:2014}. The number of unbound defects increases with activity until saturation when $\ell_{\rm a}\approx\ell_{\rm n}$, as the reduction of the nematic order parameter due to the defects compensates the activity increase \cite{Giomi:2014}. Here we overlook the problem of the onset and focus on the regime where turbulence is \emph{fully developed}, but still far from saturation: thus $\ell_{\rm n} \ll \ell_{\rm a}\ll L$. 

\revision{Experimentally, $\ell_{\rm a}$ depends on the microscopic details of the system as well as the abundance of the biochemical fuel powering the active stresses. For instance, in the microtuble-kinesin suspension shown in Fig. \ref{fig:snapshots}A, $\ell_{\rm a}\approx 100$ $\mu$m (i.e. the typical length scale associated with bending deformations), while the microtubles themselves (hence $\ell_{\rm n}$) are approximatively 1.5 $\mu$m in length \cite{Sanchez:2012}. Similar $\ell_{\rm a}$ values have been probed in experiments with actomyosin motility assays, using filaments of approximatively 5 $\mu$m in length \cite{Schaller:2013}. The latter is also the typical length of the {\em Bacillus subtilis} cells used in Ref. \cite{Wensink:2012} to investigate bacterial turbulence, while in this case $\ell_{\rm a} \approx 10$ $\mu$m, which is, thus, much closer to the lower bound of the range of length scales analyzed here.}   

\begin{figure}[t]
\centering
\includegraphics[width=1\columnwidth]{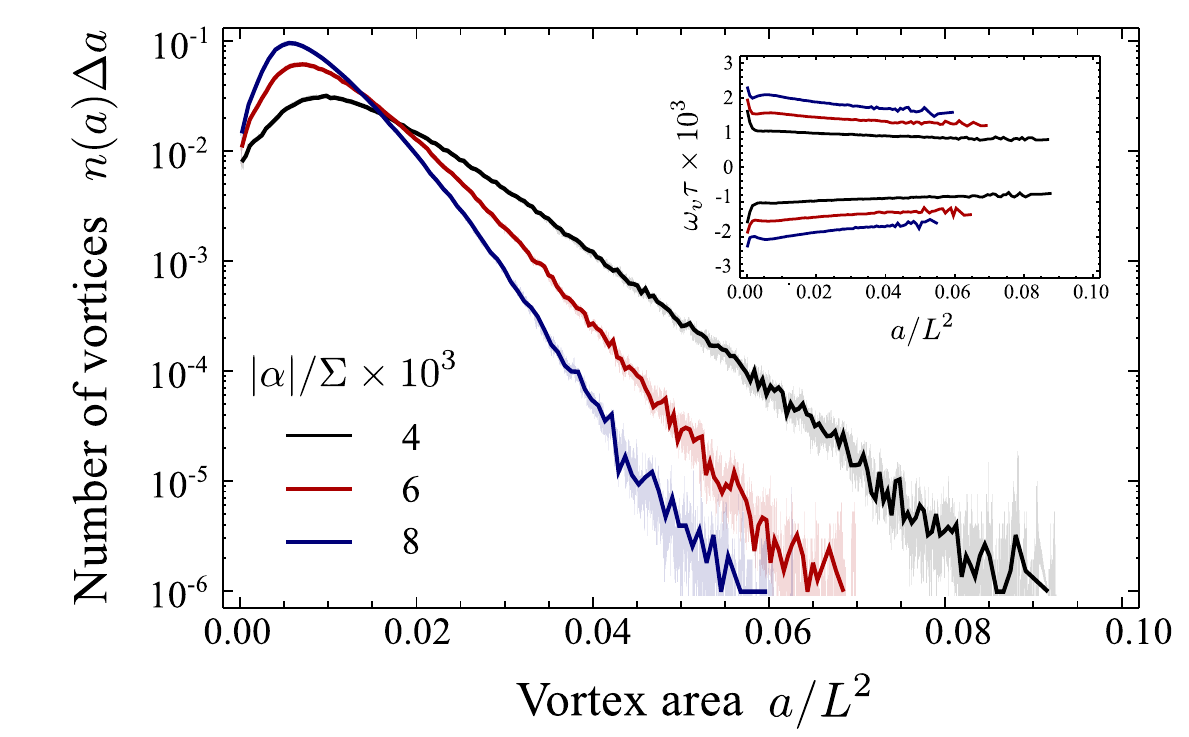}	
\caption{\label{fig:n_of_a}Number of vortices $n(a)\Delta a$ (with $\Delta a/L^2=1.5 \times 10^{-5}$) with area in between $a$ and $\Delta a$ as a function of $a$, obtained, in {\em extensile} systems, from a numerical integration of Eqs. \eqref{eq:hydrodynamics} for various $\alpha$ values. The shaded regions surrounding the curves correspond to the statistical error obtained from five simulations with different (disordered) initial conditions. The data show a prominent exponential distribution in the range $a_{\min}<a<L$, with $a_{\min}$ the area of the smallest {\em active} vortex. Inset: Average vorticity of an individual vortex as a function of its area.}
\end{figure}

Fig. \ref{fig:snapshots}B shows the typical structure of the turbulent flow arising from Eqs. \eqref{eq:hydrodynamics} for large (negative) values of the active stress $\alpha$. The velocity field appears to be decomposed in vortices of various sizes and shapes, while the director field is highly distorted by the presence of several $\pm 1/2$ disclination pairs (Fig. \ref{fig:snapshots}C). 

In order to demonstrate the multiscale structure of the flow, I measure the distribution of the vortex area. Calling $a$ the area of a vortex, this can be described by a density function $n(a)$, such that $dN=da\,n(a)$ is the total number of vortices of area in between $a$ and $a+da$. The area of a vortex can be measured from the numerical data by introducing the so-called Okubo-Weiss field \cite{Benzi:1988,Weiss:1991} $\mathcal{Q}=(\partial_{xy}^{2}\psi)^{2}-(\partial_{x}^{2}\psi)(\partial_{y}^{2}\psi)$, with $\psi$ the stream function, such that $v_{x}=\partial_{y}\psi$ and $v_{y}=-\partial_{x}\psi$. The quantity $\mathcal{Q}$ is related to the Lyapunov exponent of a tracer particle advected by the flow: where $\mathcal{Q}>0$, the distance between two initially close particles will diverge exponentially in time, while for $\mathcal{Q}<0$ the trajectories will remain close. Thus coherent regions in the flow are defined as regions in which $\mathcal{Q}<0$ \cite{Benzi:1988}. The Okubo-Weiss field has been widely used for analyzing atmospheric and oceanic circulations and realized as an important quantity to characterize two-dimensional flows \cite{Isern-Fontanet:2003,Chelton:2007}. To identify the vortex cores, on the other hand, one can calculate the angle the velocity field rotates in one loop around each cell of the computational grid \cite{Huterer:2005}. If the cell contains the core of a vortex, this angle is equal to $2\pi$, regardless of whether the vortex is left- or right-handed. The combination of these two criteria allows to formulate the following vortex detection algorithm: {\em 1)} from the velocity field $\bm{v}$ the cores of the vortices are initially located; {\em 2)} the area of a vortex is then defined as the area of the region surrounding a vortex core where $\mathcal{Q}<0$. Fig. \ref{fig:snapshots}D shows the vortices detected by this method. %Different criteria are obviously possible. For instance one could identify a vortex as the region surrounding a vortex core where the vorticity $\omega=\partial_{x}v_{y}-\partial_{y}v_{x}$ exceeds a given threshold \cite{Benzi:1988}. This procedure, however, would introduce a dependence on the chosen vorticity threshold. In general, any criterion is expected to introduce a systematic error in the calculation of the areas. Such a systematic error, however, affects the entire vortex population and therefore does not introduce any bias in the distribution $n(a)$.  

Fig. $\ref{fig:n_of_a}$ shows the density function $n(a)$ versus the vortex area $a$ obtained from a numerical integration of Eqs. \eqref{eq:hydrodynamics}. The data show a prominent exponential distribution of the form: 
\begin{equation}\label{eq:vortex_distribution}
n(a)=\frac{N}{Z}\,\exp(-a/a^{*})\;,\qquad a_{\min}<a<a_{\max}\;.	
\end{equation}
where $a_{\min}$ and $a_{\max} \sim L^{2}$ are, respectively, the minimal and maximal area of an {\em active} vortex and $a^{*}$ a suitable scale parameter. By analogy with the inertial range in classic turbulence, hereafter, we will refer to this range as the {\em active range}. Here by {\em active} vortex we mean a vortex resulting directly from mechanical work performed by the active stresses. Beside active vortices, other vortices might form due to the strong shear in the space between active vortices. These secondary vortices are expected to lie outside the active range, thus where $a<a_{\min}$. The quantities $N=\int_{a_{\min}}^{a_{\max}} da\,n(a)$ and $Z=\int_{a_{\min}}^{a_{\max}}da\,\exp(-a/a^{*})$ in Eq. \eqref{eq:vortex_distribution} represent respectively the total number of active vortices and a normalization constant. In the Conclusions I will speculate about the physical origin of this exponential distribution. The vortices mean vorticity $\omega_{v}=(1/a)\int_{\rm vortex}d^{2}r\,\omega(\bm{r})$ is shown in the inset of Fig. \ref{fig:n_of_a} as a function of $a$. Unlike $n(a)$, $\omega_{v}$ remains roughly constant across the scales and shows some dependence only for large activity values, where the mean vortex size is substantially smaller than the size of the system (see the blue line in the inset of Fig. \ref{fig:n_of_a}). 

\begin{figure}[t]
\centering
\includegraphics[width=1\columnwidth]{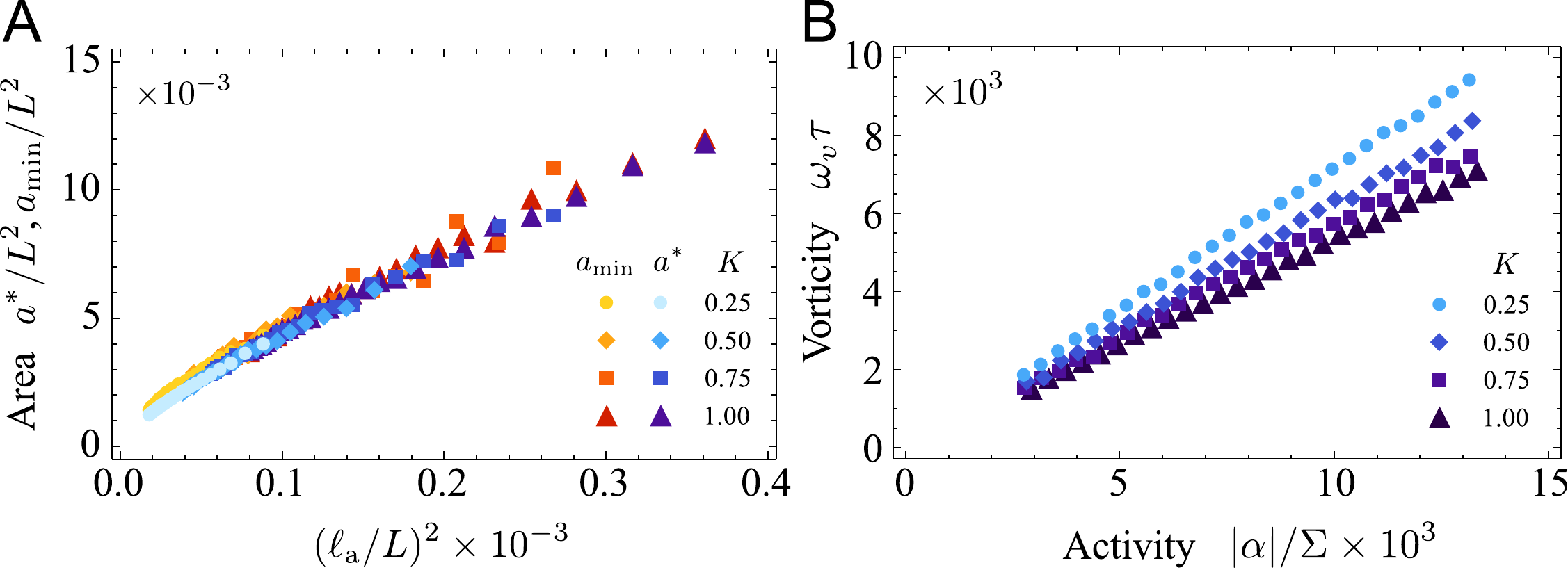}	
\caption{\label{fig:a_min_a_star}({A}) The areas $a_{\min}$ (red tones) and $a^{*}$ (blue tones) appearing in the vortex probability distribution Eq. \eqref{eq:vortex_distribution} for various activity and Frank constant $K$ values. The collapse of the data demonstrates that $a_{\min} \approx a^{*}\sim\ell_{\rm a}^{2}$. ({B}) Vortices mean vorticity $\omega_{v}$ versus activity for various $K$ values. As expected, $\omega_{v}$ grows linearly with activity, with a prefactor weakly dependent on the Frank constant.}
\end{figure}

As activity is increased, the vortices become smaller and faster as indicated by the dependence of $a_{\min}$, $a^{*}$ and $\omega_{v}$ on $\alpha$. Being $\ell_{\rm a}$ the only length scale associated with activity, intuitively one could expect that $a_{\min}\approx a^{*} \sim \ell_{\rm a}^{2}$. Analogously, the balance of active and viscous stresses over the scale of a vortex, suggests that $\omega_{v}\sim\alpha/\eta$. These expectations are confirmed from the numerical data shown in Fig. \ref{fig:a_min_a_star}.  

\revision{It is useful to recall that the microtubules-kinesin suspensions studied in \cite{Sanchez:2012,Keber:2014} consist of an active nematic monolayer at the interface of a three-dimensional bulk fluid. As was investigated in a classic paper by Stone and Ajdari \cite{Stone:1998}, the frictional damping exerted by the surrounding fluid dissipates momentum through a force of the form $\bm{f}_{\rm fri}=-\xi\bm{v}$ in Eq. (\ref{eq:hydrodynamics}a). Such a frictional interaction removes kinetic energy from the flow at scales $\ell_{\rm fri}=\sqrt{\eta/\xi}$ and is expected to have no effect on the global properties of the flow as long as $\ell_{\rm fri} \gg \ell_{\rm a}$.}

\subsection{\label{sec:flow_statistics}Statistical geometry of the flow}

\begin{figure}[t]
\centering
\includegraphics[width=1\columnwidth]{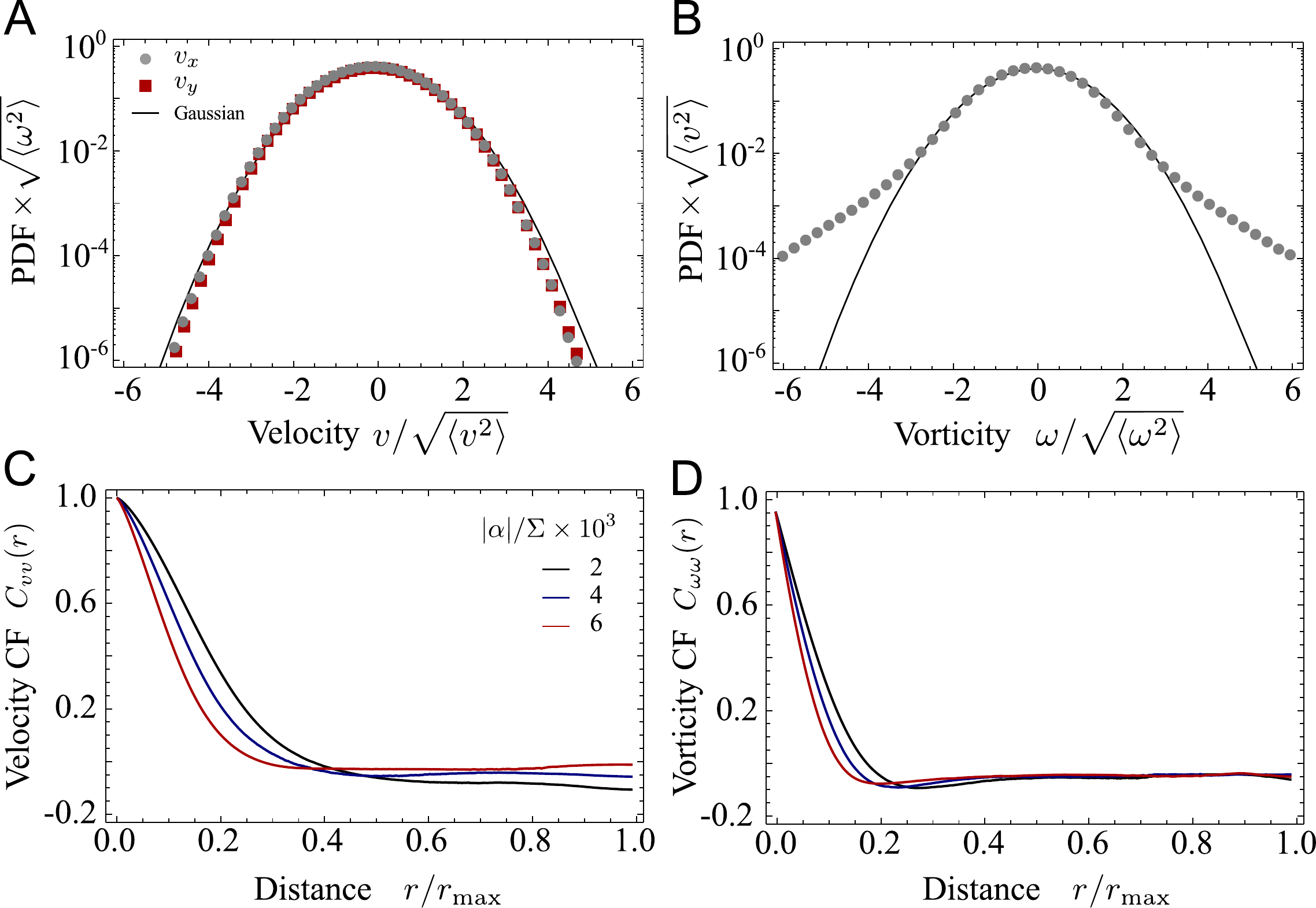}	
\caption{\label{fig:pdf}Probability distribution function of the velocity components ({A}) and vorticity ({B}). All the data are normalized by their corresponding standard deviation. The black solid line represents a unit-variance Gaussian function: $f(x)=1/\sqrt{2\pi}\,\exp(-x^{2}/2)$. Velocity ({C}) and vorticity ({D}) correlation functions for various activity values. The distance is normalized by $r_{\max}=\sqrt{2}/2\,L$, corresponding to the maximal distance between two points on a periodic square of size $L$.} 
\end{figure}

The multiscale organization and the exponential distribution of the vortex areas have striking consequences on the overall statistical properties of the flow. From a gross application of the central limit theorem we could expect the velocity components to be Gaussianly distributed. The numerical data shown in Fig. \ref{fig:pdf}A support this expectation. As in classic high Reynolds number turbulence, on the other hand, vorticity and, in general, any function of the velocity gradients do not obey the Gaussian distribution due to the spatial correlation introduced by the derivatives \cite{Firsch:1995}. In this particular case, the vorticity probability density function (PDF) exhibits a visible deviation from Gaussianity along the tails (see Fig. \ref{fig:pdf}C). 

Fig. \ref{fig:pdf}C-D show the normalized velocity and vorticity correlation functions: $C_{vv}(r)=\langle \bm{v}(0)\cdot\bm{v}(\bm{r}) \rangle/\langle |\bm{v}^{2}(0)|\rangle$ and $C_{\omega\omega}(r)=\langle \omega(0)\omega(\bm{r}) \rangle/\langle \omega^{2}(0)\rangle$, where the angular brackets $\langle\,\cdot\,\rangle$ indicate an average over space and time. These quantities have played a central role in the study of active turbulence starting from the experimental work by Sanchez {\em et al}. \cite{Sanchez:2012}. In this work it was argued that, after rescaling by the mean-squared value, the correlation functions no longer depend on activity, suggesting that the underlying geometrical structure of the flow is due to a passive mechanism, while activity controls only the flow average speed. This scenario found support in the numerical work of Thampi {\em et al}. \cite{Thampi:2013}, who also provided an elegant interpretation based on the creation and annihilation dynamics of topological defects. The numerical data shown in Fig. \ref{fig:pdf}, combined with that reported in the previous section, demonstrate that the geometrical structure of the flow does in fact depend on activity through the active length scale $\ell_{\rm a}$. Such a dependence, which is clearly marked by the intersection of the curves in Fig. \ref{fig:pdf}C-D, is, however, subtle and could have been missed before due to the limited activity range explored. An analytical approximation of the correlation functions $C_{vv}(r)$ and $C_{\omega\omega}(r)$ is given in Appendix \ref{sec:correlation_functions}.

\begin{figure}[t]
\centering
\includegraphics[width=1\columnwidth]{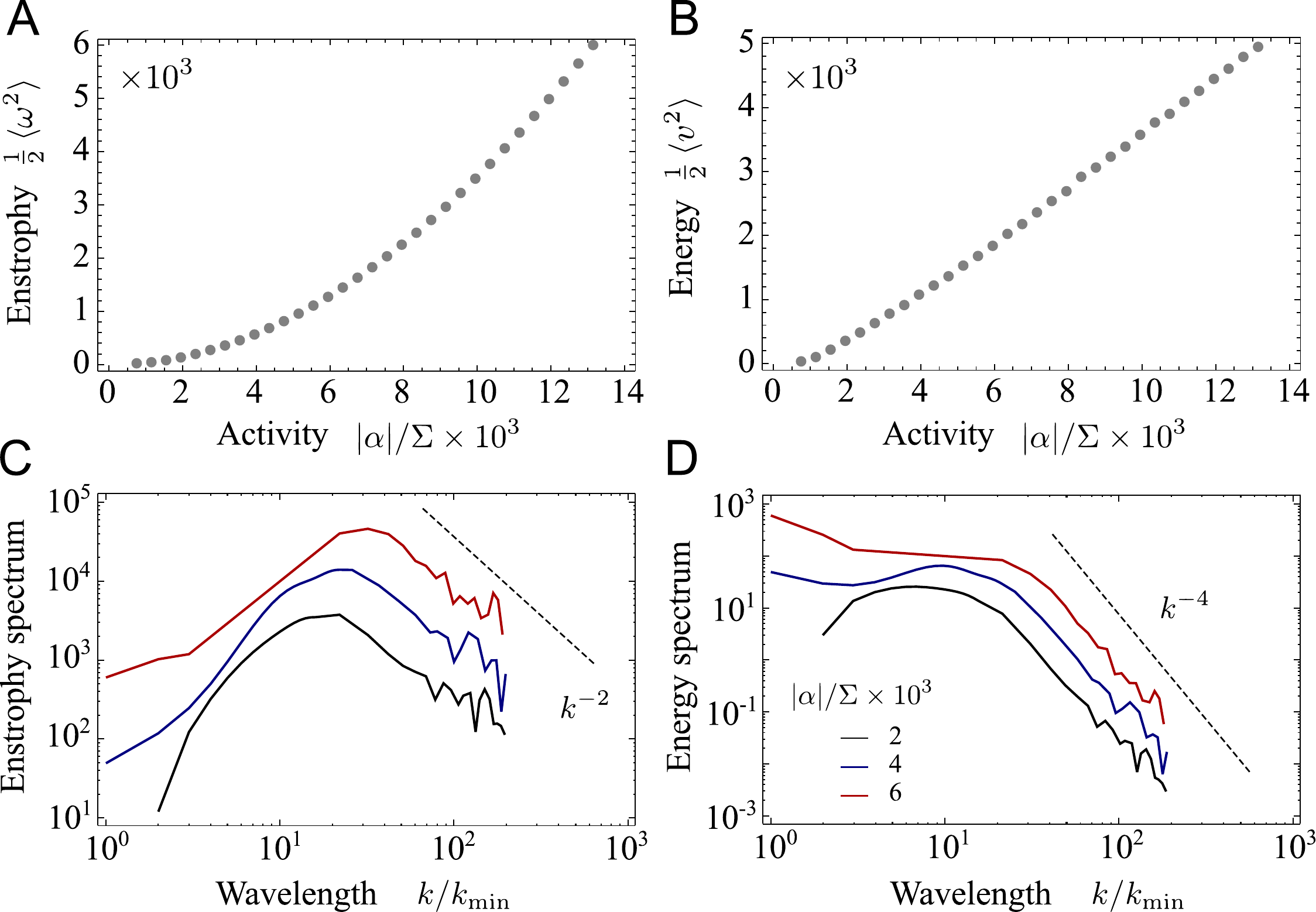}	
\caption{\label{fig:energy_enstrophy}Enstrophy ({A}) and energy ({B}) per unit area versus activity. The data show, respectively, quadratic and linear scaling. ({C}) Enstrophy and ({D}) energy spectra for three activity values. The wavelength is normalized to unity at $k_{\rm min}=2\pi/L$.} 
\end{figure}

To gain further insight about how activity affects the flow, I measured the average kinetic energy $\langle v^{2} \rangle/2$ and enstrophy $\langle \omega^{2} \rangle/2$ per unit area for varying $\alpha$ values (Fig. \ref{fig:energy_enstrophy}A-B). These exhibit, respectively, a clear linear and quadratic dependence on activity. These scaling properties can be straightforwardly understood from the geometrical picture previously described. As the vorticity is decomposed in a discrete number of vortices having $\omega_{v} \approx \alpha/\eta$, the total enstrophy can be expressed as:
\[
\Omega_{\rm tot} = \frac{1}{2}\int d^{2}r\,\omega^{2}(\bm{r}) = \frac{1}{2}\int da\,n(a)\,a\,\omega_{v}^{2}=\frac{1}{2}\,N\,\overline{a}\,\omega_{v}^{2}\;,	
\]
where $N$ is the total number of vortices and $\overline{(\,\cdot\,)}=\int da\,n(a)(\,\cdot\,)/N$ indicates the vortex ensemble average. Thus:
\begin{equation}
\frac{1}{2}\,\langle \omega^{2} \rangle = \frac{\Omega_{\rm tot}}{L^{2}} \approx \omega_{v}^{2} \sim \alpha^{2}\;,
\end{equation}
where we use the fact that $\overline{a} \approx L^{2}/N$. Analogously, the total kinetic energy of a single vortex is given by: $E_{v}(a) \approx 1/(16\pi)\,\omega_{v}^{2}a^{2}$, with the approximation becoming an equality in the case of a circular vortex. Averaging over the vortex ensemble thus gives $E_{\rm tot}=1/(16\pi)\,\omega_{v}^{2}N\,\overline{a^{2}}$, from which:
\begin{equation}
\frac{1}{2}\,\langle v^{2} \rangle \approx \omega_{v}^{2}\,\frac{\overline{a^{2}}}{\overline{a}} \sim \alpha\;.	
\end{equation} 

While changing the resolution of the vortex ensemble, activity does not affect the spectral structure of the flow. Fig. \ref{fig:energy_enstrophy}C-D, show the enstrophy $\Omega(k)$ and energy $E(k)=\Omega(k)/k^{2}$ spectra \cite{Firsch:1995} for various activity values. Analogously to what is observed in bacterial turbulence \cite{Wensink:2012}, the spectra are nonmonotonic with a peak around $k_{\rm a}=2\pi/\ell_{\rm a}$ dividing the growing regime at small $k-$values from the decay regime at large $k-$values. In the latter regime, the data show a clear power-law decay with $\Omega(k) \sim k^{-2}$ and $E(k)\sim k^{-4}$. In the next section, we illustrate the origin of these exponents in a mean-field framework.

\subsection{\label{sec:mft}Mean-field theory}

The spectral structure of turbulence in two-dimensional active nematics as well as short-scale velocity and vorticity correlation can be satisfactorily described within a mean-field approximation. This approach was introduced by Benzi {\em et al}. \cite{Benzi:1988,Benzi:1992} to account for the emergence of self-similar coherent structures in two-dimensional decaying turbulence and can be extended to the non-self-similiar case discussed here. We consider a two-dimensional flow whose vorticity field can be decomposed in a discrete number of vortices of radius $R_{i}$ and vorticity $\omega_{i}(r)=\omega_{v,i}f(r/R_{i})$, with $r$ the distance from the vortex center and $\omega_{v,i}$ a constant. Then: $\omega(\bm{r}) =\sum_{i}\omega_{v,i}f(|\bm{r}-\bm{r}_{i}|/R_{i})$, where $\bm{r}_{i}$ is the position of the $i-$th vortex center. The power spectrum of the function $\omega(\bm{r})$ can then be expressed as:
\begin{equation}
|\hat{\omega}(k)|^{2} = \sum_{ij} e^{-i\bm{k}\cdot(\bm{r}_{i}-\bm{r}_{j})}\omega_{v,i}\omega_{v,j}\,R_{i}^{2}R_{j}^{2}\hat{F}(kR_{i})\hat{F}(kR_{j})\;.
\end{equation}
where $\hat{F}(kR)=1/(2\pi)\int_{0}^{\infty} d\xi\,\xi f(\xi)J_{0}(\kappa R\,\xi)$, with $J_{0}$ a Bessel function of the first kind, is a dimensionless vortex structure factor (see Appendix \ref{sec:form_factor}). Now, if we neglect the spatial correlation between the vortices only the diagonal terms in the sum survive upon averaging. Then: 
\begin{align}\label{eq:power_spectrum}
\langle |\omega(k)|^{2}\rangle 
&= \sum_{i}\omega_{v,i}^{2}R_{i}^{4}\hat{F}^{2}(\kappa R_{i}) \notag \\
&\approx \int dR\,n(R)\,\omega_{v}^{2}(R) R^{4} \hat{F}^{2}(\kappa R)\;,
\end{align}
where we have replaced the summation with an integral over the vortex population. Finally, the enstrophy spectral density, can be calculated from the vorticity power spectrum as:  $\Omega(k) = 4\pi^{3}k\,\langle |\omega(k)|^{2} \rangle$ (see Appendix \ref{sec:spectra}). 	

Now, the distribution function $n(R)$ can be obtained from Eq. \eqref{eq:vortex_distribution} upon setting $a=\pi R^{2}$, so that $n(R)=|da/dR|n(a)$. Note that assuming the vortices to have circular shape is not generally correct as it is not guaranteed that the {\em ansatz} used to parametrize the vorticity field would hold in general. Nonetheless, based on the existence of a single characteristic length scale $\ell_{\rm a}$, one could hope that both approximations would affect the accuracy of the calculation only through irrelevant prefactors. Next, using the fact that $\omega_{v}$ does not depend on $R$ and replacing $n(R)$ into Eq. \eqref{eq:power_spectrum} yields, after some algebraic manipulations:
\begin{equation}\label{eq:enstrophy_spectrum1}
\Omega(\kappa) = \frac{8\pi^{4} \omega_{v}^{2} N}{Z}\,\left(\frac{R^{*}}{\kappa}\right)^{5} \int d\xi\,\xi^{5}e^{-\left(\frac{\xi}{\kappa}\right)^{2}}\hat{F}^{2}(\xi)\;,	
\end{equation}
where we have set $R^{*}=\sqrt{a^{*}/\pi}\sim\ell_{\rm a}$ and $\kappa=kR^{*}$. Now, consistent with the previous assumption about the shape of a vortex, we can choose $f(r/R)=1$ for $r/R\le 1$ and $f(r/R)=0$ otherwise, then the structure factor can be easily calculated in the form $\hat{F}(\xi) = J_{1}(\xi)/(2\pi\xi)$ (see Appendix \ref{sec:form_factor}). Using this in Eq. \eqref{eq:enstrophy_spectrum1} and extending for simplicity the integration to the whole positive real axis yields:
\begin{equation}\label{eq:enstrophy_spectrum2}
\Omega(\kappa) = C\kappa e^{-\frac{\kappa^{2}}{2}}\left[I_{0}\left(\frac{\kappa^{2}}{2}\right)-I_{1}\left(\frac{\kappa^{2}}{2}\right)\right]\;,
\end{equation}
where $I_{0}$ and $I_{1}$ are modified Bessel functions of the first kind \cite{Abramowitz:1972} and $C=\pi^{2}\omega_{v}^{2}N R^{*\,5}/(2Z)$ is a quantity independent on $\kappa$. The spectral structure of the turbulent flow is encoded in the asymptotic behavior of the function in Eq. \eqref{eq:enstrophy_spectrum2}. For $\kappa\gg 1$, $\Omega(\kappa)\sim\kappa^{-2}$  (see Appendix \ref{sec:spectra}) in perfect agreement with the numerical data. The energy spectral density can be calculated straightforwardly from $\Omega(\kappa)$, this yields $E(\kappa)\sim \kappa^{-4}$. For $\kappa\approx 0$, on the other hand, Eq. \eqref{eq:enstrophy_spectrum2} yields $\Omega(\kappa)\sim \kappa$ and $E(\kappa)\sim\kappa^{-1}$. These predictions are very difficult to compare with the numerical data as they refer to the narrow range of the spectrum (i.e. $k_{\min}<k<10k_{\min}$, with $k_{\min}=2\pi/L$) preceding the crossover region.

\begin{figure}[t]
\centering
\includegraphics[width=1\columnwidth]{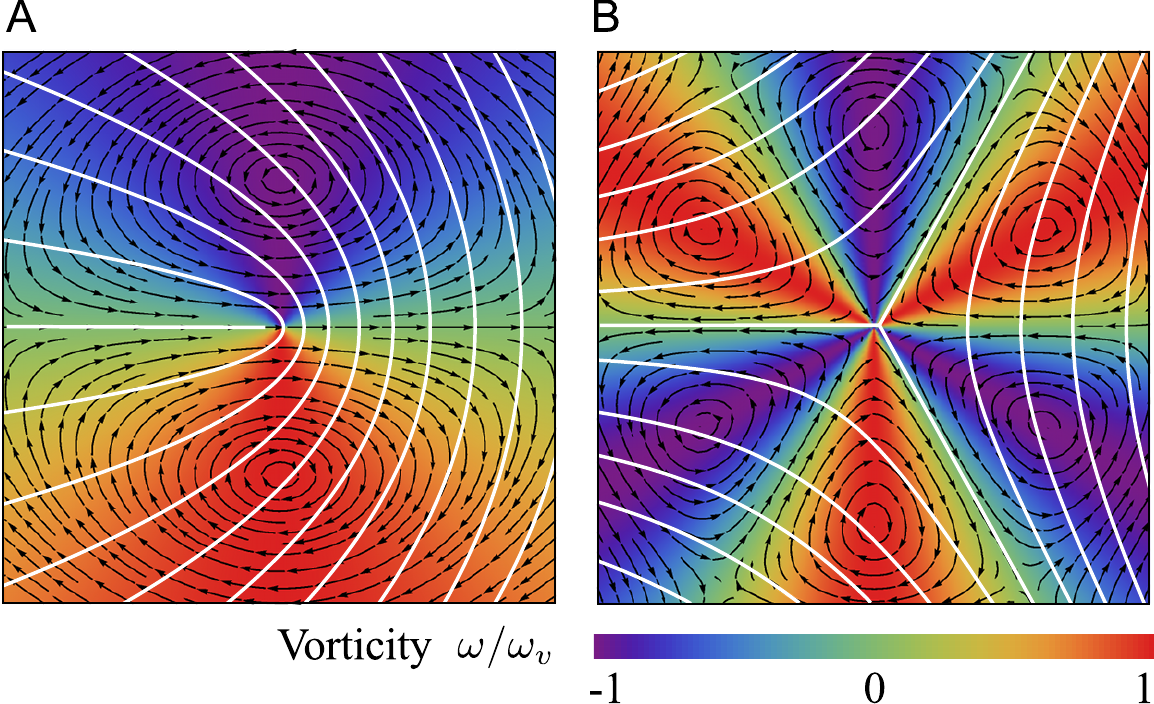}	
\caption{\label{fig:defect_flow}The flow field generated by a $+1/2$ ({A}) and $-1/2$ disclination ({B}). The white lines indicate the orientation of the director field. The black arrows correspond to the flow velocity while the background color indicates the vorticity. The flow is obtained from an analytical solution of the incompressible Stokes equation in the presence of a body force $\bm{f}^{\pm}=\nabla\cdot(\alpha\bm{Q}^{\pm})$ and $\bm{Q}^{\pm}$ the nematic tensor associated with a $\pm 1/2$ disclination \cite{Giomi:2014}.} 
\end{figure}

From the spectra $\Omega(k)$ and $E(k)$, the correlation functions $C_{vv}(r)$ and $C_{\omega\omega}(r)$ can be easily determined (see Appendix \ref{sec:correlation_functions}). Because of the mean-field approximation, however, the accuracy of this calculation is limited to the range $0<r<R^{*}$, where the spatial correlation between vortices is negligible.

\subsection{Topological structure of active turbulence}

\begin{figure}[t]
\centering
\includegraphics[width=1\columnwidth]{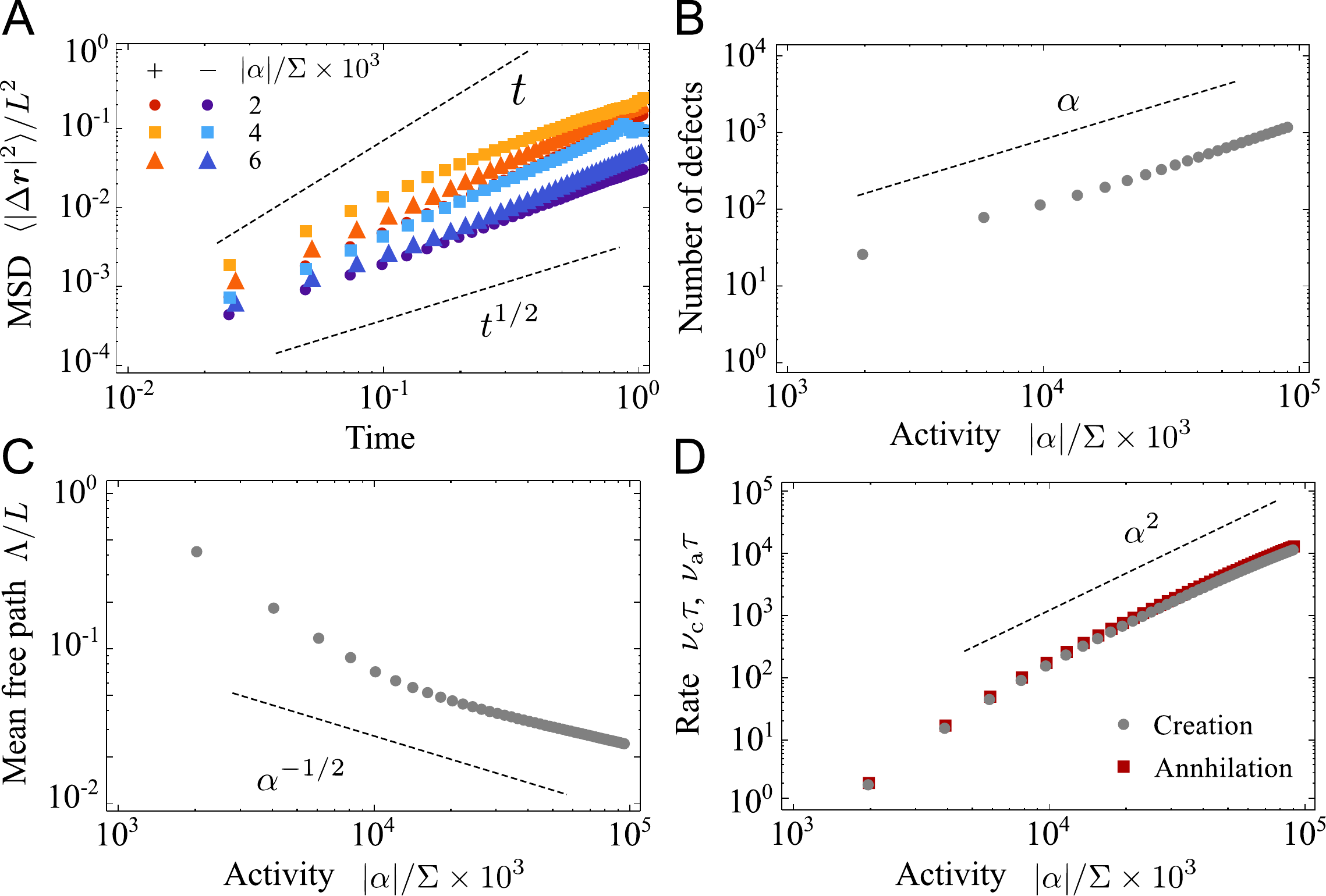}	
\caption{\label{fig:defect_stat}({A}) Mean-squared displacement of $+1/2$ (red tones) and $-1/2$ (blue tones) disclinations versus time for various activity values. Number ({B}), mean free path ({C}) and rates of creation and annihilation ({D}) of $\pm 1/2$ disclinations versus activity.} 
\end{figure}

As I mentioned in the Introduction, the geometry of the flow field is strictly connected with the topological structure of the nematic phase \cite{Sanchez:2012,Thampi:2013,Thampi:2014a,Thampi:2014b,Giomi:2013b}. As it was stressed in Ref. \cite{Giomi:2014}, the configuration of the nematic director in the neighborhood of a $\pm 1/2$ disclination determines the local vortex structure (Fig. \ref{fig:defect_flow}). Topological defects serve then as a template for the turbulent flow, which in turn advects the defects themselves leading to chaotic mixing. To provide a quantitative description of the defect chaotic dynamics I measure the mean-squared displacement (MSD) of $\pm 1/2$ disclinations as a function of time (Fig. \ref{fig:defect_stat}A). For both positively and negatively charged defects this shows a substantially diffusive behavior, with a slight superdiffusive trend in the short time dynamics of $+1/2$ disclinations, due to the self-propulsion provided by the self-induced dipolar flow (Fig. \ref{fig:defect_flow}A and Ref. \cite{Giomi:2014}). 

The total number of topological defects $N_{\rm d}$ is evidently proportional to the number of vortices. Thus $N_{\rm d} \sim L^{2}/\,\overline{a} \sim \alpha$, consistently with what we find numerically (see Fig. \ref{fig:defect_stat}B). As already mentioned, this linear growth in the defect population tends to saturate when the active length scale $\ell_{\rm a}$ approaches the defects core radius, proportional to $\ell_{\rm n}$ (not shown here) \cite{Giomi:2014}. Analogously, the defect mean-free path (Fig. \ref{fig:defect_stat}C) is $\Lambda\sim\sqrt{L^{2}/N_{\rm d}}\sim 1/\sqrt{\alpha}$. 

Fig. \ref{fig:defect_stat}D, shows the defect creating and annihilation rates $\nu_{\rm c}$ and $\nu_{\rm a}$ versus activity (in units of the viscous time scale $\tau$). For large activity values these exhibit a quadratic dependence on $\alpha$: $\nu_{\rm c}\approx\nu_{\rm a}\sim\alpha^{2}$. This behavior can be understood by noticing that topological defects moves predominantly along the edge of the vortices at approximatively constant angular velocity $\omega_{v}\approx \alpha/\eta$. During this circulation, they might approach an oppositely charged defect and annihilate. The elastic, Coulomb-like, attraction between oppositely charged defects takes over only when these have become very close to each other, thus the typical time scale of annihilation, $t_{\rm a}$, is predominantly dictated by the active circulation: $t_{\rm a}\sim 1/\omega_{v}$. From this we might expect that $\nu_{\rm a} \approx N_{\rm d}/t_{\rm a} \sim \alpha^{2}$. Once turbulence reaches a steady state, the defects creation and annihilation balance, hence $\nu_{\rm c}\approx\nu_{\rm a}$. 

\revision{The interplay between defects and vortices illustrated here for two-dimensional active nematics is both remarkable and unique, due to the asymmetric structure of semi-integer disclinations and, simultaneously, the absence of vortex stretching in two-dimensional fluids \cite{Firsch:1995}. In three-dimensional active nematics, for instance, disclination lines will give rise to tubular vortices according to the same mechanism illustrated here for the two-dimensional case. Because of vortex stretching, however, these vortices will tend to lengthen with a consequent redistribution of energy toward the small scale. Whether the effect of this redistribution will be only to bias the function $n(a)$ toward small $a$ values or more dramatic is, at the moment, impossible to predict. In active polar liquid crystals, on the other hand, the active stress associated with a $+1$ disclination does not give rise to a flow, due to the $O(2)$ symmetry of this configuration. The proliferation of $+1$ defects is then expected to hinder turbulence rather than fueling it. While this property, evidently, does not prevent low Reynolds number turbulence from developing in active polar liquid crystals, we can expect it to affect the statistics of the vortices and therefore the spectral properties of the turbulent flow.}

\section{Discussion and Conclusions}

In this article I have report a thorough numerical and analytical investigation of low Reynolds number turbulence in two-dimensional active nematics. Spectacular experimental realizations of this system are found in cytoskeletal fluids of microtubules and kinesin at the water-oil interface \cite{Sanchez:2012} or incapsulated in a lipid vesicle \cite{Keber:2014}. For large enough activity values (corresponding to high concentrations of motors or adenosine triphosaphate in cytoskeletal fluids), these systems are known to develop a chaotic spontaneous flow reminiscent of turbulence in viscous fluids (see Fig. \ref{fig:snapshots}A). 

Here I demonstrate that, as for inertial turbulence, low Reynolds number turbulence in active fluids is in fact a multiscale phenomenon characterized by the appearance of vortices spanning a range of length scales. Within this {\em active range} the vortex areas follow the exponential distribution, \revision{whose characteristic length scale $\ell_{\rm a}$ is set by the balance between active and elastic stresses}. This peculiar geometrical structure of the flow leaves a strong signature on all the relevant physical observables. The mean kinetic energy, for instance, scales linearly with activity (and not quadratically as one could have naively expected from a comparison with the laminar case, where $v\sim \alpha L/\eta$) because the vortices become smaller as activity is increased. Furthermore, the enstrophy and energy spectra scale as $k^{-2}$ and $k^{-4}$, rispectively, thus in net contrast with two-dimensional inertial turbulence \cite{Firsch:1995}. The statistics of the vortices, finally, completely determines that of the defects (and vice versa) making possible the formulation of various scaling relations amenable to experimental scrutiny.

While some questions have been answered in this work, others remain open. How do energy and enstrophy flow across the scales? In two-dimensional inertial turbulence it is well known that enstrophy flows toward the small scale where it is eventually dissipated, while energy flows toward the large scale where it is either dissipated by frictional interactions with the wall or condensed in large coherent structures \cite{Firsch:1995}. In complex fluids, on the other hand, kinetic energy can be converted into elastic energy and be dissipated or stored via mechanisms that do not require cascading. While the existence of an {\em active range} does imply that of energy and enstrophy flux across scales, the organization of such a flux remains unknown. 

Another important question, which I deliberately saved for the end, concerns the origin of the exponential distribution of the vortex areas. Perhaps the most natural explanation appeals to the interpretation of the vortex population as an equilibrium ensemble, subject to the laws of statistical mechanics. This reasoning is not new in two-dimensional turbulence, but goes back to the pioneering works of Onsager \cite{Onsager:1949} and of Joyce and Montgomery \cite{Joyce:1973,Montgomery:1974} (see  also Refs. \cite{Eyink:2006,Chorin:1997} for a review). To clarify this concept let us consider a system of $N$ active vortices having the same absolute vorticity and let $n_{i}$ be the number of vortices of area $a_{i}$, so that $\sum_{i}n_{i}=N$. A microscopic configuration is then characterized by a set of occupancy numbers $\{n_{i}\}_{i=1}^{\infty}$ and, as the vortices are indistinguishable, there are $W=N!/\prod_{i}n_{i}!$ different ways to realize the same microscopic configuration. In the limit of large $N$, corresponding to fully developed active turbulence, $W\sim e^{\mathcal{S}}$, where $\mathcal{S} = -\int da\,n(a)\log n(a)$ is an analog of the \revision{Shannon-Gibbs entropy} for the vortex ensemble. Since the vortices all have the same vorticity, a macroscopic state can be arguably identified by the their total number $N=\int da\,n(a)$ and area $\mathcal{A}=\int da\,n(a)\,a$, with $\overline{a}=\mathcal{A}/N$. The most probable $(N,\mathcal{A})-$macrostate is that maximizing the entropy $\mathcal{S}$ for fixed $N$ and $\mathcal{A}$; hence, $n(a)=N/\overline{a}\,\exp(-a/\overline{a})$. Finally, setting $\overline{a}\sim\ell_{\rm a}^{2}$ yields an expression equivalent to Eq. \eqref{eq:vortex_distribution}.

This hypothesis is inevitably naive and yet incredibly fascinating in suggesting an unexpected connection between the simplest and the most complex forms of matter.

\begin{acknowledgments}
This work is supported by The Netherlands Organization for Scientific Research (NWO/OCW). I am grateful to Luca Heltai, Cristina Marchetti, Sumesh Thampi, Vincenzo Vitelli and Julia Yeomans for several useful conversations and especially indebted with Stephen DeCamp and Zvonimir Dogic for the experimental image of Fig. \ref{fig:snapshots}A as well as all the wonderful work that inspired this research.
\end{acknowledgments}

\appendix 

\section{\label{sec:form_factor}Vortex form factor}

The dimensionless vortex structure factor $\hat{F}(kR)$, introduced in the mean-field calculation, is defined from the Fourier transform of the vorticity profile function $f(r/R)$, where $r$ is the distance from the vortex center. Thus:
\begin{align}\label{eq:structure_factor}
\hat{f}(k) 
&=\int \frac{d^{2}r}{(2\pi)^{2}}\,e^{-i\bm{k}\cdot\bm{r}}f\left(\frac{r}{R}\right) \notag \\
&= \frac{1}{2\pi}\int_{0}^{\infty} dr\,r f\left(\frac{r}{R}\right)J_{0}(kr)\;,	
\end{align}
where $J_{0}$ is a zeroth-order Bessel function of the first kind. Now, the simplest approximation of the function $f$ is evidently:
\begin{equation}
f\left(\frac{r}{R}\right)=\left\{
\begin{array}{lll}
1 & $\,$ & r/R \le 1 \\[10pt]
0 & $\,$ & r/R > 1\;,
\end{array}
\right.	
\end{equation} 
representing a circular vortex of radius $R$. Placing this into Eq. \eqref{eq:structure_factor} yields:
\begin{equation}
\hat{f}(k) 
= \frac{1}{2\pi} \int_{0}^{R} dr\,r J_{0}(kr)	
= \frac{R}{2\pi k}\,J_{1}(kR)\;.
\end{equation}
The dimensionless function $\hat{F}(kR)$ is then defined from $\hat{f}(k)=R^{2}\hat{F}(kR)$; hence:
\begin{equation}
\hat{F}(kR)=\frac{1}{2\pi k R}\,J_{1}(kR)\;.
\end{equation}

\section{\label{sec:spectra}Asymptotic behavior of the spectral densities}

The expression of the enstrophy spectrum, as obtained within the mean-field approximation, is given by:
\begin{equation}\label{eq:enstrophy_spectrum}
\Omega(\kappa)=C\kappa e^{-\frac{\kappa^{2}}{2}}\left[I_{0}\left(\frac{\kappa^{2}}{2}\right)-I_{1}\left(\frac{\kappa^{2}}{2}\right)\right]\;.	
\end{equation}
Now, small $\kappa$ limit can be easily determined by considering that, for $\kappa\ll 1$, $I_{0}(\kappa^{2}/2)\approx\exp(-\kappa^{2}/2)\approx 1$ and $I_{1}(\kappa^{2}/2)\approx 0$. Therefore:
\begin{equation}
\Omega(\kappa)\sim \kappa\;, \qquad \kappa \ll 1\;.	
\end{equation}
To calculate  the large $\kappa$ limit we can use the following asymptotic expansion of the modified Bessel function:
\begin{equation}
I_{\nu}(x) \approx \frac{e^{x}}{\sqrt{2\pi x}}\left(1+\frac{1-4\nu^{2}}{8x}+\cdots\right)\;.	
\end{equation}
From this we obtain:
%\begin{align*}
%I_{0}\left(\frac{\kappa^{2}}{2}\right) &\approx \frac{e^{\frac{\kappa^{2}}{2}}}{\sqrt{\pi}\kappa}\left(1+\frac{1}{4\kappa^{2}}\right)\;,\\[10pt]
%I_{1}\left(\frac{\kappa^{2}}{2}\right) &\approx \frac{e^{\frac{\kappa^{2}}{2}}}{\sqrt{\pi}\kappa}\left(1-\frac{3}{4\kappa^{2}}\right)\;,
%\end{align*}
%then:
\begin{equation}
I_{0}\left(\frac{\kappa^{2}}{2}\right)-I_{1}\left(\frac{\kappa^{2}}{2}\right) \approx \frac{e^{\frac{\kappa^{2}}{2}}}{\sqrt{\pi}\kappa^{3}}\;.
\end{equation}
The exponential term exactly cancels that in Eq. \eqref{eq:enstrophy_spectrum} resulting in a simple power-law behavior:
\begin{equation}
\Omega(\kappa)\sim \kappa^{-2}\;,\qquad \kappa\gg 1\;.	
\end{equation}
The asymptotic behavior of the energy spectrum follows directly from this by virtue of the relation $\Omega(k)=k^{2}E(k)$. 

\section{\label{sec:correlation_functions}Correlation functions}

\begin{figure}[t]
\centering
\includegraphics[width=0.8\columnwidth]{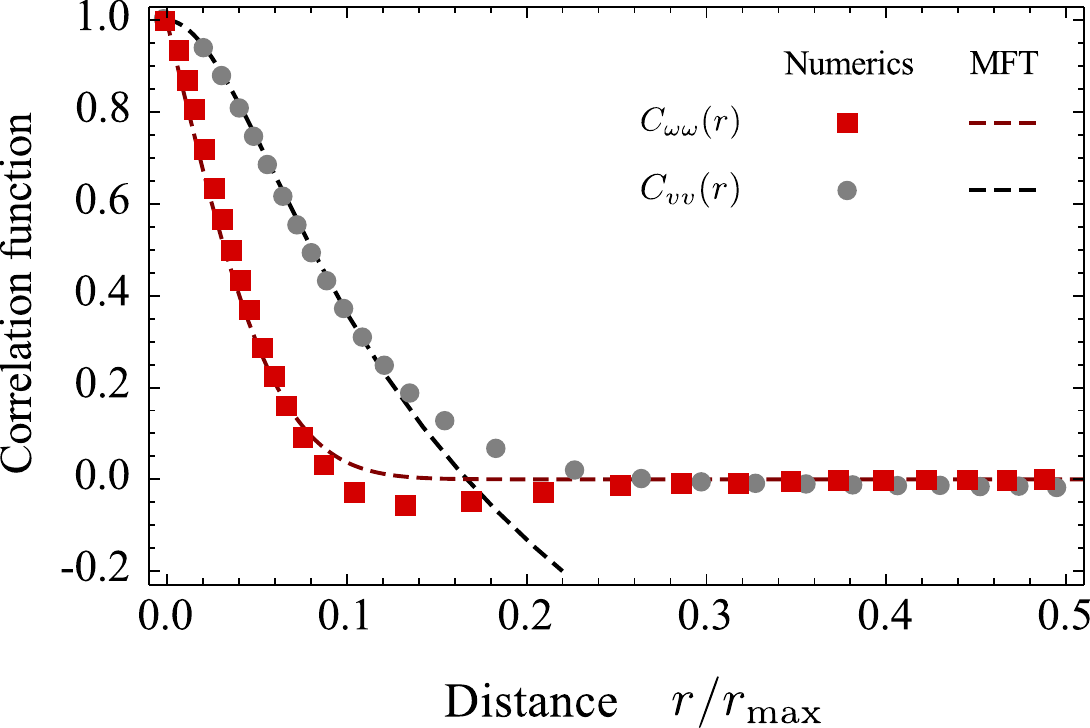}
\caption{\label{fig:correlation_functions}Normalized vorticity (red dots) and velocity (gray dots) correlation functions obtained from a numerical integration of Eqs. \eqref{eq:hydrodynamics}. The theoretical curves (MFT) correspond to Eqs. \eqref{eq:vorticity_correlation_function} and \eqref{eq:velocity_correlation_function}, with $R^{*}=0.034\,r_{\max}$ obtained from a fit of the data. An extrapolation of $R^{*}$ from the scale parameter $a^{*}$, as defined in Eq. \eqref{eq:vortex_distribution}, can be obtained by setting $R^{*}=\sqrt{a^{*}/\pi}$, this gives $R^{*}=0.054\,r_{\max}$ slightly larger than the value obtained from a fit of the correlation function. This slight discrepancy is presumably due to the inevitable systematic error in the calculation of the vortex area from the Okubo-Weiss field as well as the circular approximation of the vortex shape.}	
\end{figure}

The spectral densities $E(k)$ and $\Omega(k)$ and the correlation functions $\langle \bm{v}(0)\cdot\bm{v}(r) \rangle$ and $\langle \omega(0)\omega({\bm r}) \rangle$ are related by the Weiner-Kinchin theorem \cite{Firsch:1995}. This implies that: 
\begin{equation}\label{eq:wiener_kinchin}
\Omega(k) = \frac{1}{2}\,\Delta_{d}\,k^{d-1}\mathcal{F}\{\langle \omega(0)\omega(\bm{r}) \rangle\}\;,	
\end{equation}
where $\Delta_{d}$ is the $d-$dimensional solid angle and $\mathcal{F}$ denotes Fourier transformation. An equivalent expression holds for the vorticity spectrum and, in general, for the power spectrum of any random field given its two-point correlation function. For the special case of a two-dimensional vorticity field with azimuthal symmetry, Eq. \eqref{eq:wiener_kinchin} yields:
\begin{equation}\label{eq:enstrophy_integral}
\Omega(k) = \frac{1}{2} \int_{0}^{\infty} dr\, kr J_{0}(kr) \langle \omega(0)\omega(\bm{r}) \rangle\;.
\end{equation}
If the spectrum is known, Eq. \eqref{eq:enstrophy_integral} can be inverted to obtain the vorticity correlation function $C_{\omega\omega}(r)=\langle \omega(0)\omega(\bm{r}) \rangle /\langle |\omega(0)|^{2} \rangle$. This yields:
\begin{equation}\label{eq:vorticity_correlation_function}
C_{\omega\omega}(r) = \frac{2}{\langle \omega^{2} \rangle} \int_{0}^{\infty} dk\,J_{0}(kr)\Omega(k)\;,	
\end{equation}
where $\langle \omega^{2} \rangle/2=\int_{0}^{\infty}dk\,\Omega(k)$ is the mean enstrophy per unit area. The same expression holds for the velocity correlation function upon replacing $\Omega(k)$ with $E(k)$ and the normalization factor with the mean energy per unit area: $\langle v^{2} \rangle/2 = \int_{0}^{\infty}dk\,E(k)$. 

Now, placing the expression for $\Omega(k)$ given into Eq. \eqref{eq:enstrophy_spectrum} in \eqref{eq:vorticity_correlation_function} yields:
\begin{equation}
C_{\omega\omega}(r) = \erfc\left(\frac{r}{2R^{*}}\right)\;,	
\end{equation}
where $\erfc(x)=1-\erf(x)$ is the complementary error function \cite{Abramowitz:1972} while $2R^{*}\sim\ell_{\rm a}$ represent the mean diameter of an active vortex. 

\begin{figure}[t]
\centering
\includegraphics[width=0.8\columnwidth]{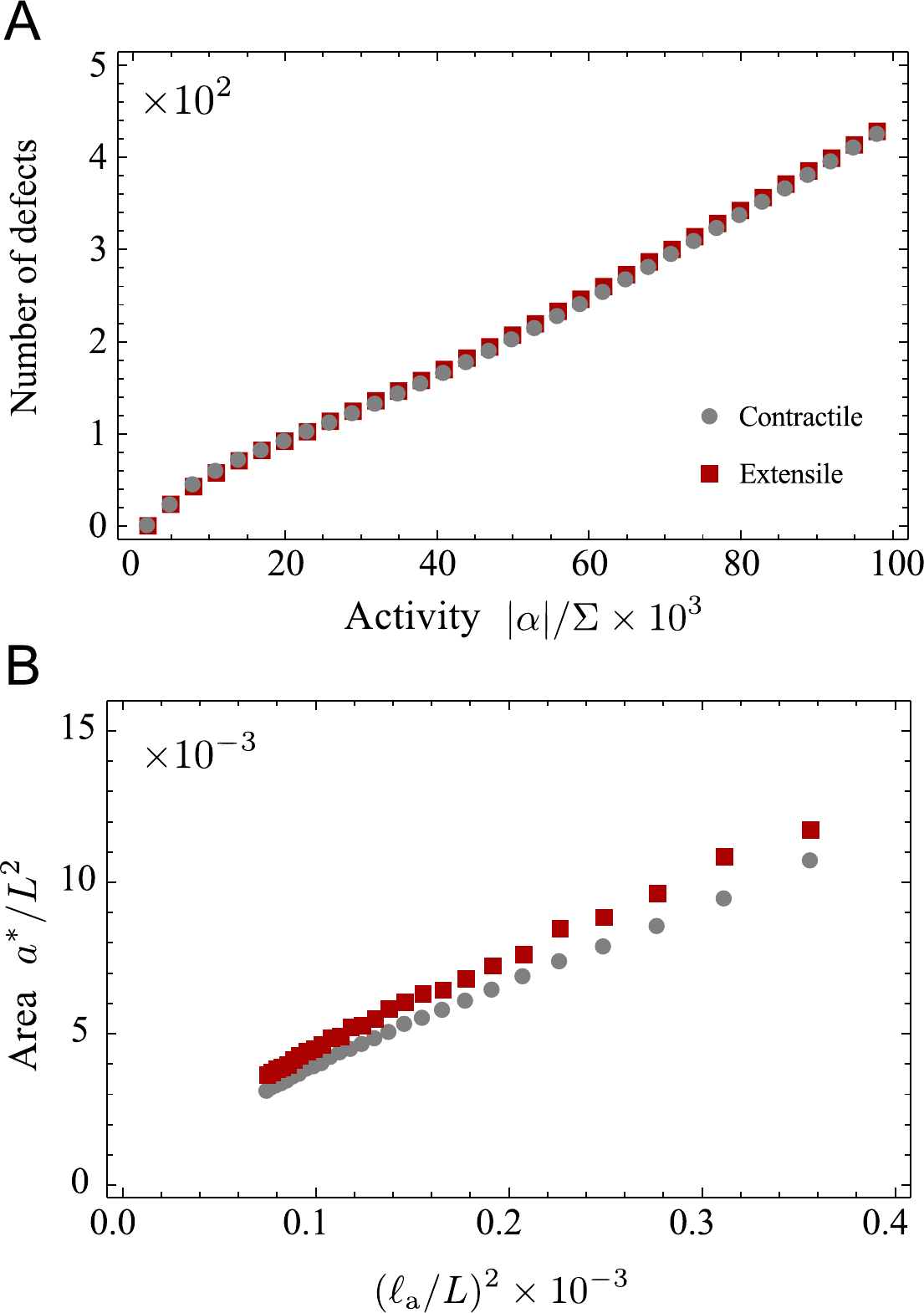}
\caption{\label{fig:contractile_vs_extensile}({A}) Number of defects and scale factor $a^{*}$ versus activity for contractile (gray dots) and extensile (red dots) systems. ({B}) The scale factor $a^{*}$ appearing in the vortex area distribution $n(a)$ versus the active length scale $\ell_{\rm a}=\sqrt{K/|\alpha|}$ for for contractile (gray dots) and extensile (red dots) systems.}	
\end{figure}

The simple algebraic relation between energy and enstrophy spectra, translates to real space into the following differential relation between the velocity and vorticity correlation functions \cite{Chorin:1997}: 
\begin{equation}
\nabla^{2}\langle \bm{v}(0)\cdot\bm{v}(\bm{r})\rangle = -\langle \omega(0)\omega(\bm{r}) \rangle\;.
\end{equation}
For an azimuthally symmetric function, this implies:
\begin{equation}\label{eq:velocity_correlation_function}
C_{vv}(r) = 1-\frac{\langle \omega^{2} \rangle}{\langle v^{2} \rangle} \int_{0}^{r}dr'\,\frac{h(r')}{r'}\;,
\end{equation}
where $h(r)=\int_{0}^{r}dr'\,r'C_{\omega\omega}(r)$. Fig. \ref{fig:correlation_functions} shows a comparison between the normalized velocity and vorticity correlation functions obtained from a numerical integration of Eqs. \eqref{eq:hydrodynamics} and their mean-field approximations given in Eqs. \eqref{eq:vorticity_correlation_function} and \eqref{eq:velocity_correlation_function}. For small distances, the agreement is remarkable. This, however, breaks down for $r\gg R^{*}$, where the spatial correlation between neighboring vortices (which is neglected in the mean-field framework) becomes crucial. For instance, $C_{\omega\omega}(r)$ becomes negative when $r$ is larger than the average vortex diameter, due to the fact that a given central vortex is surrounded by vortices of opposite vorticity (see the red dots in Fig. \ref{fig:correlation_functions}). This feature is clearly absent in the mean-field calculation and the resulting correlation function decays without sign changes (solid black line in Fig. \ref{fig:correlation_functions}).  

\section{\label{sec:extensile_vs_contractile}Extensile versus contractile}

The numerical results presented in the main text describe the case of extensile active nematics ($\alpha<0$), such as the suspensions of microtubule bundles and kinesin pioneered by Sanchez {\em et al}. \cite{Sanchez:2012} and recently employed by Keber {\em et al}. in the fabrication of active vesicles \cite{Keber:2014}. The behavior of contractile active nematics ($\alpha>0$) is nearly identical to the extensile case. For a given activity magnitude $|\alpha|$ the average number of defects in contractile and extensile systems (therefore the spatial organization of the flow) is essentially the same (Fig. \ref{fig:contractile_vs_extensile}A). 

The only notable difference appears to be the offset in the linear relation between $a^{*}$ and $\ell_{\rm a}^{2}$ (Fig. \ref{fig:contractile_vs_extensile}B). Such an offset is presumably due to the asymmetry between contractile and extensile systems at the onset of turbulence \cite{Giomi:2014}, which is in turn related to the asymmetry in the linear instability of the quiescent state. I remand the reader to Refs. \cite{Edwards:2009,Giomi:2014} for a detailed explanation.


\begin{thebibliography}{99}

\bibitem{Ramaswamy:2010}
S. Ramaswamy,
{\em The mechanics and statistics of active matter},
\href{http://dx.doi.org/10.1146/annurev-conmatphys-070909-104101}{Annu. Rev. Condens. Matter Phys. {\bf 1}, 323 (2010)}
[\arxivhref{http://arxiv.org/abs/1004.1933}{arXiv:1004.1933}].

\bibitem{Vicsek:2012}
T. Vicsek, and A. Zafeiris,
{\em Collective motion},
\href{http://dx.doi.org/10.1016/j.physrep.2012.03.004}{Phys. Rep. {\bf 517}, 71 (2012)}
[\arxivhref{http://arxiv.org/abs/1010.5017}{arXiv:1010.5017}].

\bibitem{Marchetti:2013}
M. C. Marchetti, J.-F. Joanny, S. Ramaswamy, T. B. Liverpool, J. Prost, M. Rao, and R. A. Simha RA,  
{\em Hydrodynamics of soft active matter},
\href{http://dx.doi.org/10.1103/RevModPhys.85.1143}{Rev. Mod. Phys. {\bf 85}, 1143 (2013)}
[\arxivhref{http://arxiv.org/abs/1207.2929}{arXiv:1207.2929}].

\bibitem{Vicsek:1995}
T. Vicsek, A. Czirok, E. Ben-Jacob, I. Cohen, and O. Shochet,
{\em Novel type of phase transition in a system of self-driven particles},
\href{http://dx.doi.org/10.1103/PhysRevLett.75.1226}{Phys. Rev. Lett. {\bf 75}, 1226 (1995)}.

\bibitem{Ballerini:2008}	
M. Ballerini, N. Cabibbo, R. Candelier, A. Cavagna, E. Cisbani, I. Giardina, V. Lecomte, A. Orlandi, G. Parisi, A. Procaccini, M. Viale, and V. Zdravkovic,
{\em Interaction ruling animal collective behavior depends on topological rather than metric distance: Evidence from a field study},
\href{http://dx.doi.org/10.1073/pnas.0711437105}{Proc. Natl. Acad. Sci. USA {\bf 105}, 1232 (2008)}
[\arxivhref{http://arxiv.org/abs/0709.1916}{arXiv:0709.1916}].

\bibitem{Giomi:2013a}
L. Giomi, N. Hawley-Weld, and L. Mahadevan,
{\em Swarming, swirling and stasis in sequestered bristle-bots},
\href{http://dx.doi.org/10.1098/rspa.2012.0637}{Proc. R. Soc. A {\bf 469}, 20120637 (2013)}
[\arxivhref{http://arxiv.org/abs/1302.5952}{arXiv:1302.5952}].

\bibitem{Dombrowski:2004}
C. Dombrowski, L. Cisneros, S. Chatkaew, R. E. Goldstein, and J. O. Kessler,
{\em Self-concentration and large-scale coherence in bacterial dynamics},
\href{http://dx.doi.org/10.1103/PhysRevLett.93.098103}{Phys. Rev. Lett. {\bf 93}, 098103 (2004)}.

\bibitem{Wogelmuth:2008}
C. W. Wogelmuth,
{\em Collective swimming and the dynamics of bacterial turbulence},
\href{http://dx.doi.org/10.1529/biophysj.107.118257}{Biophys. J. {\bf 95}, 1564 (2008)}.

\bibitem{Zhang:2010}
H. P. Zhang, A. Be’er, E.-L. Florin, H. L. Swinney, 
{\em Collective motion and density fluctuations in bacterial colonies},
\href{http://dx.doi.org/10.1073/pnas.1001651107}{Proc. Natl. Acad. Sci. U.S.A. {\bf 107}, 13626 (2010)}.

\bibitem{Wensink:2012}
H. H. Wensink, J. Dunkel, S. Heidenreich, K. Drescher, R. E. Goldstein, H. L\"owen, and J. M. Yeomans,
{\em Meso-scale turbulence in living fluids},
\href{http://dx.doi.org/10.1073/pnas.1202032109}{Proc. Natl. Acad. Sci. USA {\bf 109}, 14308 (2012)}
[\arxivhref{http://arxiv.org/abs/1208.4239v1}{arXiv:1208.4239}].

\bibitem{Dunkel:2013}
J. Dunkel, S. Heidenreich, K. Drescher, H. H. Wensink, M. B\"ar, and R. E. Goldstein,
{\em Fluid dynamics of bacterial turbulence}
\href{http://dx.doi.org/10.1103/PhysRevLett.110.228102}{Phys. Rev. Lett. {\bf 110}, 228102 (2013)}
[\arxivhref{http://arxiv.org/abs/1302.5277}{arXiv:1302.5277}].

\bibitem{Bricard:2013}
A. Bricard, J. B. Caussin, N. Desreumaux, O. Dauchot, and D. Bartolo,
{\em Emergence of macroscopic directed motion in populations of motile colloids},
\href{http://dx.doi.org/10.1038/nature12673}{Nature {\bf 503}, 95 (2013)}
[\arxivhref{http://arxiv.org/abs/1311.2017}{arXiv:1311.2017}].

\bibitem{Palacci:2013}
J. Palacci, S. Sacanna, A. Preska-Steinberg, D. J. Pine, and P. M. Chaikin, 
{\em Living Crystals of Light Activated Colloidal Surfers},
\href{http://dx.doi.org/10.1126/science.1230020}{Science {\bf 339}, 936 (2013)}.

\bibitem{Kruse:2004}
K. Kruse, J.-F. Joanny, F. J\"ulicher, J. Prost, and K. Sekimoto,
{\em Asters, vortices, and rotating spirals in active gels of polar filaments},
\href{http://dx.doi.org/10.1103/PhysRevLett.92.078101} {Phys. Rev. Lett. {\bf 92}, 078101 (2004)}.

\bibitem{Voituriez:2005}
R. Voituriez, J.-F. Joanny, and J. Prost,
{\em Spontaneous flow transition in active polar gels},
\href{http://dx.doi.org/10.1209/epl/i2004-10501-2}{\emph{Europhys. Lett.} {\bf 70}, 404 (2005)}
[\arxivhref{http://arxiv.org/abs/q-bio.SC/0503022}{arXiv:q-bio/0503022v1}].

\bibitem{Kruse:2005}
K. Kruse, J.-F. Joanny,  F. J\"ulicher, J. Prost, and K. Sekimoto,
{\em Generic theory of active polar gels: a paradigm for cytoskeletal dynamics}
\href{http://dx.doi.org/10.1140/epje/e2005-00002-5}{\emph{Eur. Phys. J. E} {\bf 16}, 5 (2005)}
[\arxivhref{http://arxiv.org/abs/physics/0406058v1}{arXiv:physics/0406058}].

\bibitem{Schaller:2013}
V. Schaller, and A. R. Bausch,
{\em Topological defects and density fluctuations in collectively moving systems},
\href{http://dx.doi.org/10.1073/pnas.1215368110}{Proc. Nat. Acad. Sci. U.S.A. {\bf 110}, 4488 (2013)}.

\bibitem{Sanchez:2012}
T. Sanchez, D. N. Chen, S. J. DeCamp, M. Heymann, Z. Dogic,
{\em Spontaneous motion in hierarchically assembled active matter},
\href{http://dx.doi.org/10.1038/nature11591} {Nature {\bf 491}, 431 (2012)}
[\arxivhref{http://arxiv.org/abs/1301.1122v1}{arXiv:1301.1122}].

\bibitem{Keber:2014}
F. C. Keber, E. Loiseau, T. Sanchez, S. J. DeCamp, L. Giomi, M. J. Bowick, M. C. Marchetti, Z. Dogic, and A. R. Bausch,
{\em Topology and dynamics of active nematic vesicles},
\href{http://dx.doi.org/10.1126/science.1254784}{\emph{Science} {\bf 345}, 1135 (2014)}.

\bibitem{Zhou:2014}
S. Zhou, A. Sokolov, O. D. Lavrentovich, and I. S. Aranson,
{\em Living liquid crystals},
\href{http://dx.doi.org/10.1073/pnas.1321926111}{Proc. Nat. Acad. Sci. U.S.A. {\bf  111}, 1265 (2014)}
[\arxivhref{http://arxiv.org/abs/1312.5359v1}{arXiv:1312.5359}].

\bibitem{Thampi:2013}
S. P. Thampi, R. Golestanian, and J. M. Yeomans,
{\em Velocity correlations in an active nematic},
\href{http://dx.doi.org/10.1103/PhysRevLett.111.118101}{Phys. Rev. Lett. {\bf 111}, 118101 (2013)}
[\arxivhref{http://arxiv.org/abs/1302.6732v2}{arXiv:1302.6732}].

\bibitem{Thampi:2014a}
S. P. Thampi, R. Golestanian, and J. M. Yeomans,
{\em Instabilities and topological defects in active nematics},
\href{http://dx.doi.org/10.1209/0295-5075/105/18001}{Europhys. Lett. {\bf 105}, 18001 (2014)}
[\arxivhref{http://arxiv.org/abs/1312.4836v1}{arXiv:1312.4836}].

\bibitem{Thampi:2014b}
S. P. Thampi, R. Golestanian, and J. M. Yeomans,
{\em Vorticity, defects and correlations in active turbulence},
\href{http://dx.doi.org/10.1098/rsta.2013.0366}{Phil. Trans. R. Soc. A 372, 20130366 (2014)}
[\arxivhref{http://arxiv.org/abs/1402.0715}{arXiv:1402.0715}].

\bibitem{Giomi:2013b}
L. Giomi, M. J. Bowick, X. Ma, and M. C. Marchetti,
{\em Defect annihilation and proliferation in active nematics},
\href{http://dx.doi.org/10.1103/PhysRevLett.110.228101}{Phys. Rev. Lett. {\bf 110}, 228101 (2013)}
[\arxivhref{http://arxiv.org/abs/1303.4720v2}{arXiv:1303.4720}].

\bibitem{Giomi:2014}
L. Giomi, M. J. Bowick, P. Mishra, R. Sknepnek, and M. C. Marchetti,
{\em Defect dynamics in active nematics},
\href{http://dx.doi.org/10.1098/rsta.2013.0365}{Phil. Trans. R. Soc. A 372, 20130365 (2014)}
[\arxivhref{http://arxiv.org/abs/1403.5254}{arXiv:1403.5254}].

\bibitem{Gao:2014}
T. Gao, R. Blackwell, M. A. Glaser, M. D. Betterton, M. J. Shelley,
{\em A multiscale active nematic theory of microtubule/motor-protein assemblies},
\href{http://dx.doi.org/10.1103/PhysRevLett.114.048101}{Phys. Rev. Lett. 114, 048101 (2015)}
[\arxivhref{http://arxiv.org/abs/1401.8059}{arXiv:1401.8059}].

\bibitem{Shi:2013}
X. Shi, and Y. Ma,
{\em Topological structure dynamics revealing collective evolution in active nematics},
\href{http://dx.doi.org/10.1038/ncomms4013}{Nat Commun. {\bf 4}, 3013 (2014)}.

\bibitem{Marenduzzo:2007}
D. Marenduzzo, E. Orlandini, M. E. Cates, and J. M. Yeomans,
{\em Steady-state hydrodynamic instabilities of active liquid crystals: Hybrid lattice Boltzmann simulations},
\href{http://dx.doi.org/10.1103/PhysRevE.76.031921}{Phys. Rev. E {\bf 76}, 031921 (2007)}
[\arxivhref{http://arxiv.org/abs/0708.2062v1}{arXiv:0708.2062}].

\bibitem{Giomi:2012}
L. Giomi, L. Mahadevan, B. Chakraborty, M. F. Hagan,
{\em Banding, excitability and chaos in active nematic suspensions},
\href{http://dx.doi.org/10.1088/0951-7715/25/8/2245} {Nonlinearity {\bf 25}, 2245 (2012)}
[\arxivhref{http://arxiv.org/abs/1110.4338v1}{arXiv:1110.4338}].

\bibitem{Liverpool:2008}
T. B. Liverpool, and M. C. Marchetti,
{\em Hydrodynamics and rheology of active polar filaments},
\href{http://dx.doi.org/10.1007/978-0-387-73050-9_7}{in \emph{Cell Motility}, P. Lenz editor, 177-206 (Springer, New York)}
[\arxivhref{http://arxiv.org/abs/q-bio/0703029v1}{arXiv:q-bio/0703029}].

\bibitem{Lau:2009}
A. W. C. Lau, and T. C. Lubensky,
{\em Fluctuating hydrodynamics and microrheology of a dilute suspension of swimming bacteria},
\href{http://dx.doi.org/10.1103/PhysRevE.80.011917} {Phys. Rev. E {\bf 80}, 011917 (2009)}.

\bibitem{DeGennes:1993}
de Gennes PG, Prost J.
\emph{The Physics of Liquid Crystals}: 2nd edn. 
Oxford: Oxford University Press (1993).

\bibitem{Edwards:2009}
S. A. Edwards, and J. M. Yeomans,
{\em Spontaneous flow states in active nematics: a unified picture},
\href{http://dx.doi.org/10.1209/0295-5075/85/18008}{Europhys. Lett. {\bf 85}, 18008 (2009)}
[\arxivhref{http://arxiv.org/abs/0811.3432v1}{arXiv:0811.3432}].

\bibitem{Benzi:1988}
R. Benzi, S. Patarnello, P. Santangelo,
{\em Self-similar coherent structures in two-dimensional decaying turbulence},
\href{http://dx.doi.org/10.1088/0305-4470/21/5/018} {J. Phys. A: Math. Gen. {\bf 21}, 1221 (1988)}.

\bibitem{Weiss:1991}
J. Weiss,
{\em The dynamics of enstrophy transfer in two-dimensional hydrodynamics},
\href{http://dx.doi.org/10.1016/0167-2789(91)90088-Q} {Physica D {\bf 48}, 273 (1991)}.

\bibitem{Isern-Fontanet:2003}
J. Isern-Fontanet, E. Garc\'ia-Ladona, and J. Font,
{\em Identification of marine eddies from altimetric maps},
\href{http://dx.doi.org/10.1175/1520-0426(2003)20%3C772:IOMEFA%3E2.0.CO;2} {J. Atmos. Oceanic Technol. {\bf  20}, 772 (2003)}.

\bibitem{Chelton:2007}
D. B. Chelton, M. G. Schlax, R. M. Samelson, and R. A. de Szoeke, 
{\em Global observations of large oceanic eddies},
\href{http://dx.doi.org/10.1029/2007GL030812} {Geophys. Res. Lett. {\bf 34}, L15606 (2007)}.

\bibitem{Huterer:2005}
D. Huterer, and T. Vachaspati,
{\em Distribution of singularities in the cosmic microwave background polarization},
\href{http://dx.doi.org/10.1103/PhysRevD.72.043004} {Phys. Rev. D {\bf  72}, 043004 (2005)}
[\arxivhref{http://arxiv.org/abs/astro-ph/0405474}{arXiv:astro-ph/0405474}].

\bibitem{Stone:1998}
H. A. Stone, and A. Ajdari,
{\em Hydrodynamics of particles embedded in a flat surfactant layer overlying a subphase of finite depth},
J. Fluid Mech. {\bf 369}, 151 (1998).

\bibitem{Firsch:1995}
U. Frisch,
\emph{Turbulence:the legacy of A.N. Kolmogorov},
(Cambridge University Press, Cambridge, 1995).

\bibitem{Benzi:1992}
R. Benzi, M. Colella, M. Briscolini, and P. Santangelo,
{\em A simple point vortex model for two-dimensional decaying turbulence},
\href{http://dx.doi.org/10.1063/1.858254}{Phys. Fluids A {\bf 4}, 1036 (1992)}.

\bibitem{Abramowitz:1972}
M. Abramowitz, and I. A. Stegun, 
\emph{Handbook of mathematical functions with formulas, graphs, and mathematical tables: 9th ed.} 
(Dover, New York, 1972).

\bibitem{Onsager:1949}
L. Onsager,
{\em Statistical hydrodynamics},
\href{http://dx.doi.org/10.1007/BF02780991}{\emph{Nuovo Cimento Suppl.} {\bf 6}, 279 (1949)}.

\bibitem{Joyce:1973}
G. Joyce, and D. Montgomery,
{\em Negative temperature states for the two-dimensional guiding-centre plasma},
\href{http://dx.doi.org/10.1017/S0022377800007686}{J. Plasma Phys. {\bf 10}, 107 (1973)}.

\bibitem{Montgomery:1974}
D. Montgomery, and G. Joyce,
{\em Statistical mechanics of ``negative temperature'' states},
\href{http://dx.doi.org/10.1063/1.1694856}{Phys. Fluids {\bf 17}, 1139 (1974)}.

\bibitem{Eyink:2006} 
G. L. Eyink, K. R. Sreenivasan,
{\em Onsager and the theory of hydrodynamic turbulence},
\href{http://dx.doi.org/10.1103/RevModPhys.78.87}{Rev. Mod. Phys. {\bf 78}, 87 (2006)}.

\bibitem{Chorin:1997}
A. J. Chorin, 
\emph{Vorticity and Turbulence},
(Springer, New York, 1997).

\end{thebibliography}
\end{document}